\documentclass[journal,letterpaper]{IEEEtran}
\usepackage[utf8]{inputenc}
\usepackage{babel}
\usepackage{cite}
\usepackage{amsmath,amssymb,amsfonts}
\usepackage{graphicx}
\usepackage[detect-weight=true, detect-family=true]{siunitx}
\usepackage[siunitx]{circuitikz}
\usepackage{xcolor}
\usepackage{hyperref}
\hypersetup{
    colorlinks,
    citecolor=black,
    filecolor=black,
    linkcolor=black,
    urlcolor=black
}
\usepackage{subcaption}

\usepackage{graphicx}
\graphicspath{{images}}

\begin{document}

\title{The UmboMic: A PVDF Cantilever Microphone}
\author{
Aaron J. Yeiser*, Emma F. Wawrzynek*, John Z. Zhang, Lukas Graf, Christopher I. McHugh, Ioannis Kymissis, Elizabeth S. Olson, Jeffrey H. Lang*, Hideko Heidi Nakajima*
\thanks{Submitted December 2023. This paper is partially supported by NIH Grant R01 DC016874, NSF GRFP Grant 1745302, NSF GRFP Grant 2141064, a grant from the Clo\"{e}tta Foundation, Z\"{u}rich Switzerland, and the Research Fund of the University of Basel, Switzerland.}
\thanks{A. J. Yeiser was with the MIT Electrical Engineering and Computer Science (EECS) Department, Cambridge, MA, 02139.  He is now an independent engineering consultant (e-mail: ayeiser@mit.edu).}
\thanks{E. F. Wawrzynek is with the MIT Electrical Engineering and Computer Science (EECS) Department, Cambridge, MA, 02139. (e-mail: emmafw@mit.edu).}
\thanks{J. Z. Zhang is with the MIT Mechanical Engineering Department (e-mail: johnz@mit.edu).}
\thanks{L. Graf is with the Harvard Medical School and Mass. Eye and Ear Department of OHNS (e-mail: lukas.graf@usb.ch)}
\thanks{C. I. McHugh is with the Harvard Medical School and Mass. Eye and Ear Department of OHNS (e-mail: christopher\_mchugh@meei.harvard.edu)}
\thanks{I. Kymissis is with Columbia University Department of Electrical Engineering (e-mail: johnkym@ee.columbia.edu)}
\thanks{E. S. Olson is with Columbia University Departments of OTO/HNS and Biomedical Engineering (e-mail: eao2004@cumc.columbia.edu)}
\thanks{J. H. Lang is with the MIT EECS department (e-mail: lang@mit.edu).}
\thanks{H. H. Nakajima is with the Harvard Medical School and Eaton-Peabody Laboratories at Mass. Eye and Ear Department of OHNS (e-mail: heidi\_nakajima@meei.harvard.edu).}
}
\date{April 2022}
\maketitle

\begin{abstract}
\textit{Objective:} We present the ``UmboMic," a prototype piezoelectric cantilever microphone designed for future use with totally-implantable cochlear implants.
\textit{Methods:} The UmboMic sensor is made from polyvinylidene difluoride (PVDF) because of its low Young’s modulus and biocompatibility. The sensor is designed to fit in the middle ear and measure the motion of the underside of the eardrum at the umbo. To maximize its performance, we developed a low noise charge amplifier in tandem with the UmboMic sensor. This paper presents the performance of the UmboMic sensor and amplifier in fresh cadaveric human temporal bones.
\textit{Results:} When tested in human temporal bones, the UmboMic apparatus achieves an equivalent input noise of \SI{32.3}{dB}~SPL over the frequency range \SI{100}{Hz} to \SI{7}{kHz}, good linearity, and a flat frequency response to within \SI{10}{dB} from about \SI{100}{Hz} to \SI{6}{kHz}.
\textit{Conclusion:}
These results demonstrate the feasibility of a PVDF-based microphone when paired with a low-noise amplifier. The reported UmboMic apparatus is comparable in performance to a conventional hearing aid microphone.
\textit{Significance:}
The proof-of-concept UmboMic apparatus is a promising step towards creating a totally-implantable cochlear implant. A completely internal system would enhance the quality of life of cochlear implant users. 
\end{abstract}

\section{Introduction}

\IEEEPARstart{C}{ochlear} implants are a successful neuroprosthetic that can restore hearing to people with severe sensorineural hearing loss. While partially implanted, they rely on an external hearing aid microphone that is positioned on the side of the head. The external nature of this microphone imposes many lifestyle restrictions on cochlear implant users. Patients cannot swim or play certain sports while wearing the external unit, nor can they wear it while sleeping. Additionally, an external microphone does not provide the pressure gain and sound localization cues derived from the outer ear structure.
Engineering a practical internal microphone would enable a totally-implantable cochlear implant. Although development of implantable microphones has been ongoing for years, none are currently on the market. Technical approaches range from fiber-optic vibrometery~\cite{optical} to capacitive displacement sensing~\cite{capacitive}. Two devices are currently in clinical trials: a piezoelectric sensor called the Acclaim by Envoy~\cite{envoy_acclaim}~\cite{envoy_acclaim_news} and a subcutaneous microphone called Mi2000 by MED-EL~\cite{mi2000}~\cite{mi2000_news}. There is very little information available about either device, and they remain in testing.

\begin{figure}
  \centering
  \includegraphics[width=\columnwidth]{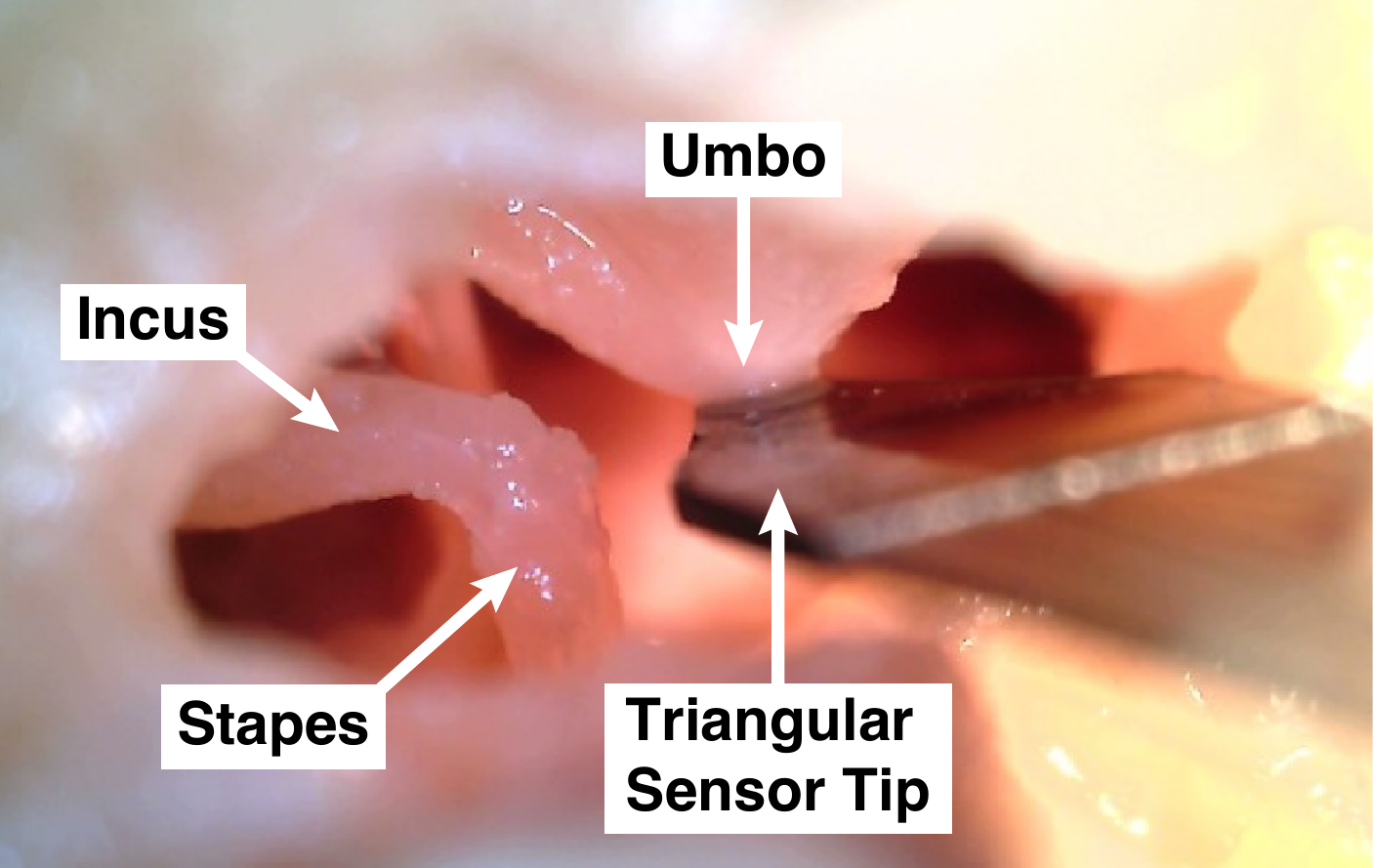}
  \caption{
    An image of a human cadaveric middle ear cavity with the UmboMic sensor inserted. The UmboMic sensor tip is touching the underside of the eardrum at the umbo. The umbo is the point where the end of the manubrium of the malleus attaches to the tympanic membrane. The malleus, incus, and stapes make up the ossicular chain.
  }
  \label{fig:middle_ear}
\end{figure}

The microphone reported here is a piezoelectric sensor paired with a charge amplifier that we call the ``UmboMic". We refer to the piezoelectric sensor as the ``UmboMic sensor" and the sensor connected to the amplifier as the ``UmboMic apparatus." The UmboMic sensor detects the motion of the umbo, which is the tip of the malleus that attaches to the conical point on the underside of the eardrum. Figure~\ref{fig:middle_ear} shows a picture of the UmboMic sensor in contact with a human umbo.
Umbo displacement is large for all auditory frequencies and mostly unidirectional in humans, making it an ideal target for sensing motion. By sensing the umbo, the UmboMic apparatus has an advantage over microphones that target other parts of the ossicular chain. For example, the Acclaim by Envoy targets the incus body, which has complex modes of motion around ~\SI{2}{kHz}.

\begin{figure*}[t]
  \centering
  \includegraphics[width=\textwidth]{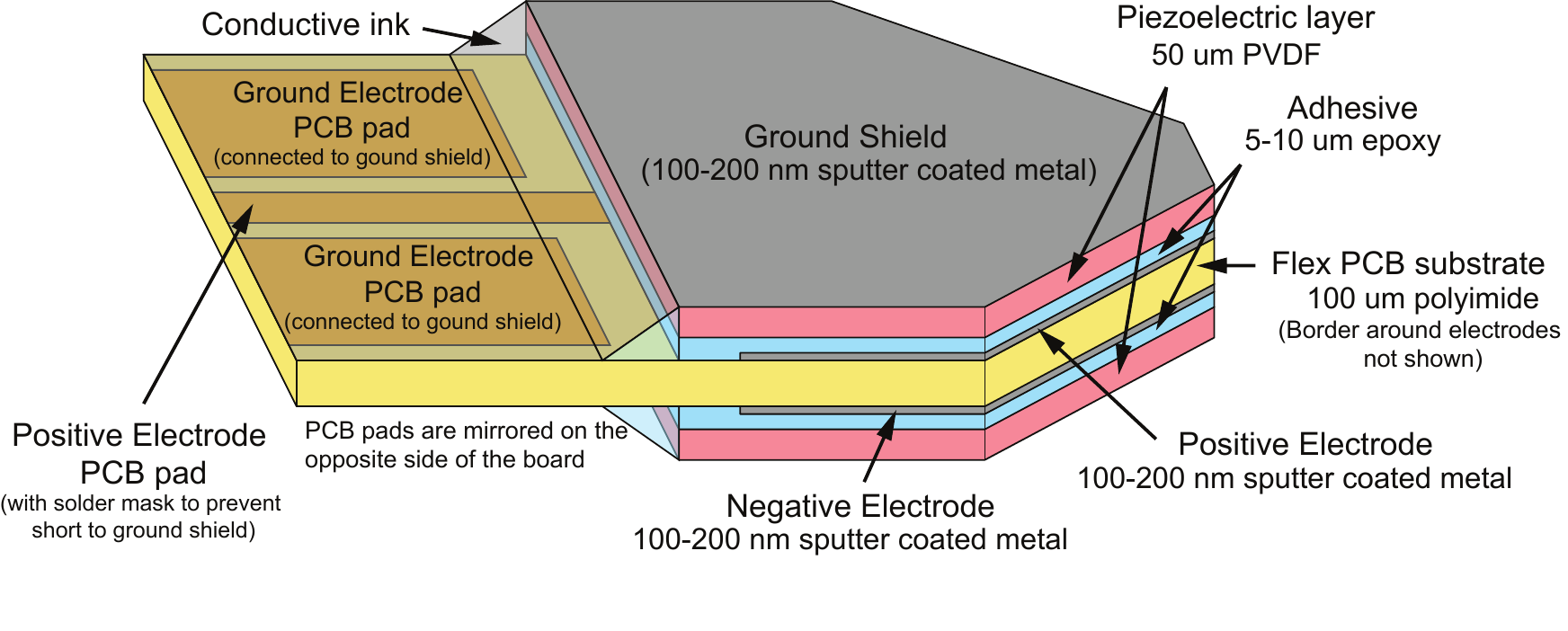}
  \caption{
    Differential PVDF cantilever diagram (not to scale).
    We construct the cantilever from two layers of PVDF sandwiching a flex PCB base.
    We sputter-coat the charge sense electrodes onto the flex PCB and capacitively couple the electrodes to the PVDF through the thin glue layer. Note that the border around the electrodes is not shown in this figure.
  }
  \label{fig:diagram}
\end{figure*}

We build the UmboMic sensor out of a thin film piezoelectric polymer called polyvinylidene difluoride (PVDF). PVDF is excellent for our application because it is highly flexible and biocompatible~\cite{PVDF}. Typically, PVDF is considered a poor choice for small-area sensors as it is less sensitive than piezoelectric ceramics. To overcome this limitation, our design relies on the differential measurement between two layers of PVDF connected to an extremely low noise amplifier to boost the signal-to-noise ratio. This paper presents a prototype PVDF sensor and an accompanying custom low-noise differential charge amplifier. The reported UmboMic apparatus exhibits high sensitivity and low noise comparable to commercially available hearing-aid microphones such as the Sonion 65GG31T~\cite{65GG31T} and the Knowles EK3103~\cite{ek3103}. In the next steps, we are advancing the microphone with fully biocompatible, decades-durable materials.

\section{Cantilever design and fabrication}
The UmboMic sensor is a triangular bimorph cantilever approximately \SI{3}{mm} wide at the base, \SI{3}{mm} long, and \SI{200}{\um} thick. The free end of the triangular tip interfaces with the umbo to sense its motion.
We design the UmboMic sensor to have a relatively uniform stress distribution in the PVDF.
The UmboMic sensor is fabricated with two layers of \SI{50}{\um} PVDF sandwiching a \SI{100}{\um} Kapton flexible printed-circuit-board (flex PCB) substrate; this construction is detailed in the following sections and in Figure~\ref{fig:diagram}.
% The charge sense electrodes are enclosed by the ground shield and insulating plastic layers of the sensor, and are capacitively coupled to the PVDF through a thin glue layer.
The use of a Kapton flex PCB as the core layer greatly simplifies attaching cables to the device.
Additionally, the PCB design allows for the ground electrode to double as a ground shield, which works in tandem with the differential sensor output to nearly eliminate electromagnetic interference.

\subsection{Designing sensor dimensions}

We use a triangular shape for the UmboMic sensor as it results in a uniform stress and charge distribution throughout the sensor tip. The triangular shape is a design commonly used with piezoelectric sensors and actuators~\cite{WOOD} as it increases the sensor’s robustness by equalizing stress concentration. A triangular shaped sensor is also practical given the anatomical limitations of the middle ear. The sensor's tapered shape allows it to slide into position without hitting the other ossicles during insertion. 

There are a few factors to consider when deciding on UmboMic sensor geometry. Our sensors must be small enough to fit through a variety of middle-ear cavity surgical entrances. However, in order to maximize the charge output of our piezoelectric sensor, we want its active surface area to be as large as possible. A larger sensor is also faster and cheaper to fabricate. We found through testing that a \SI{3}{mm} by \SI{3}{mm} triangular sensor tip fits well within the middle ear cavity of multiple cadaveric specimens, and is large enough to produce a sufficient output charge.

Further details on the UmboMic's sensor design are detailed in~\cite{Aaron_masters}.

% \section{Fabrication details}

% This section describes a replicable fabrication process, while also identifying fabrication process variations that were not as successful.
% Two of the biggest challenges to fabricating a successful cantilever was the integration of charge sense electrodes, and PVDF adhesion.
% The current design employing capacitive coupling through a very thin epoxy layer was chosen after many adhesion failures involving other adhesives and conductive tapes.

\subsection{Electrode patterning}

\begin{figure}[t]
  \centering
  \begin{subfigure}{\columnwidth}
    \includegraphics[width=\columnwidth]{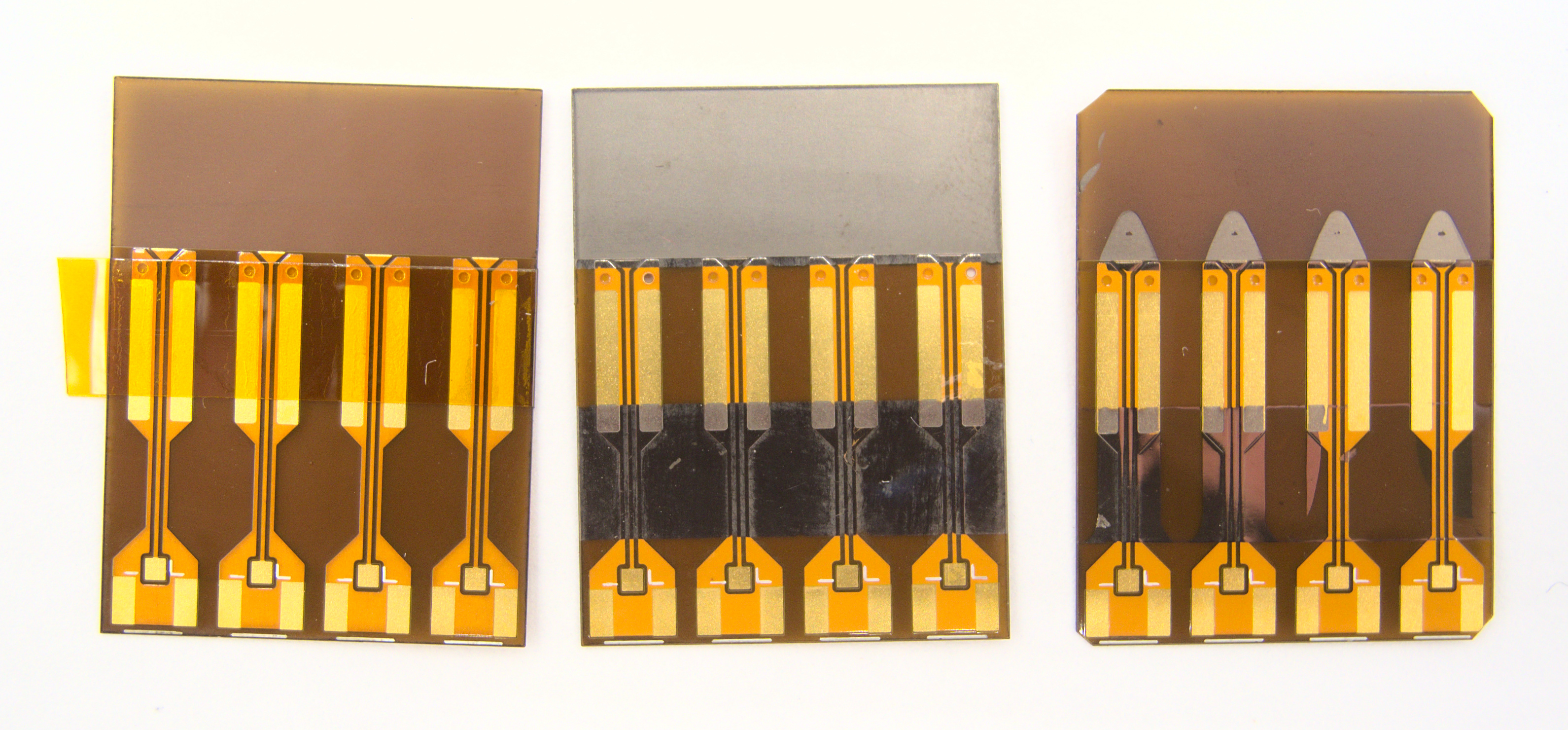}
    \caption{Left to right: We sputter-coat \SI{200}{nm} of aluminum onto a PCB and then pattern into the shape of the electrodes using contact photolithography.}
    \label{fig:fab_pattern}
  \end{subfigure}

  \begin{subfigure}{\columnwidth}
    \includegraphics[width=\columnwidth]{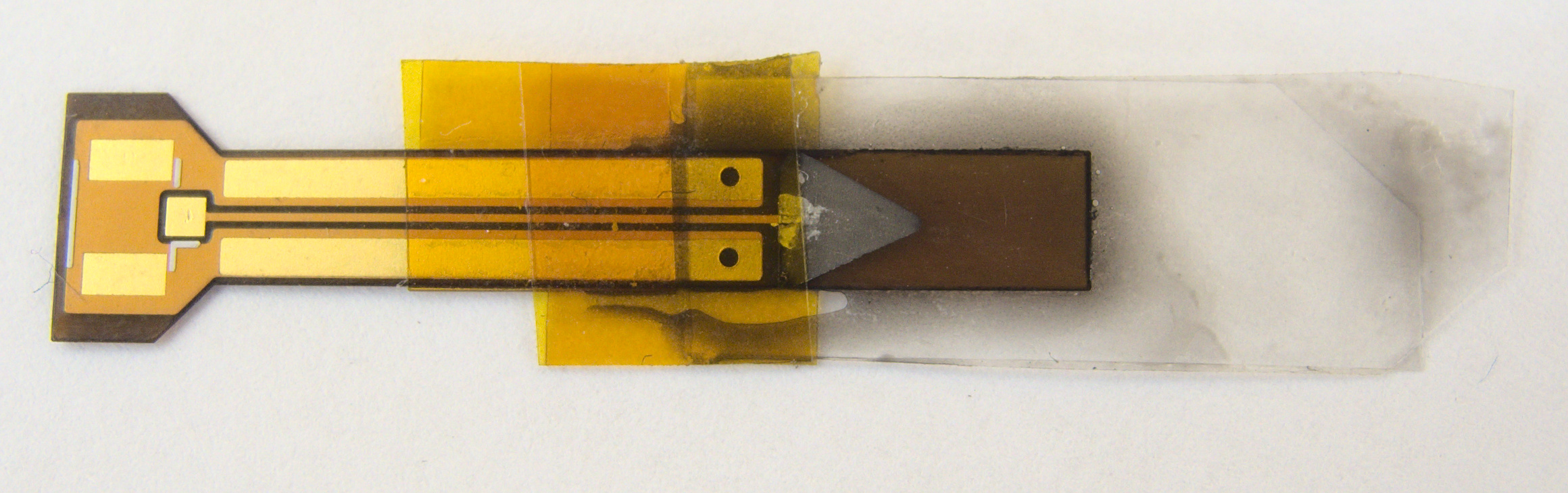}
    \caption{We apply conductive ink at the interface between the patterned electrode and the PCB electrode pattern. We then use Kapton tape to mask off areas of the flex PCB and glue the PVDF on each side with epoxy. The PVDF is the transparent film seen in the image.}
    \label{fig:fab_tape}
  \end{subfigure}

  \begin{subfigure}{\columnwidth}
    \includegraphics[width=\columnwidth]{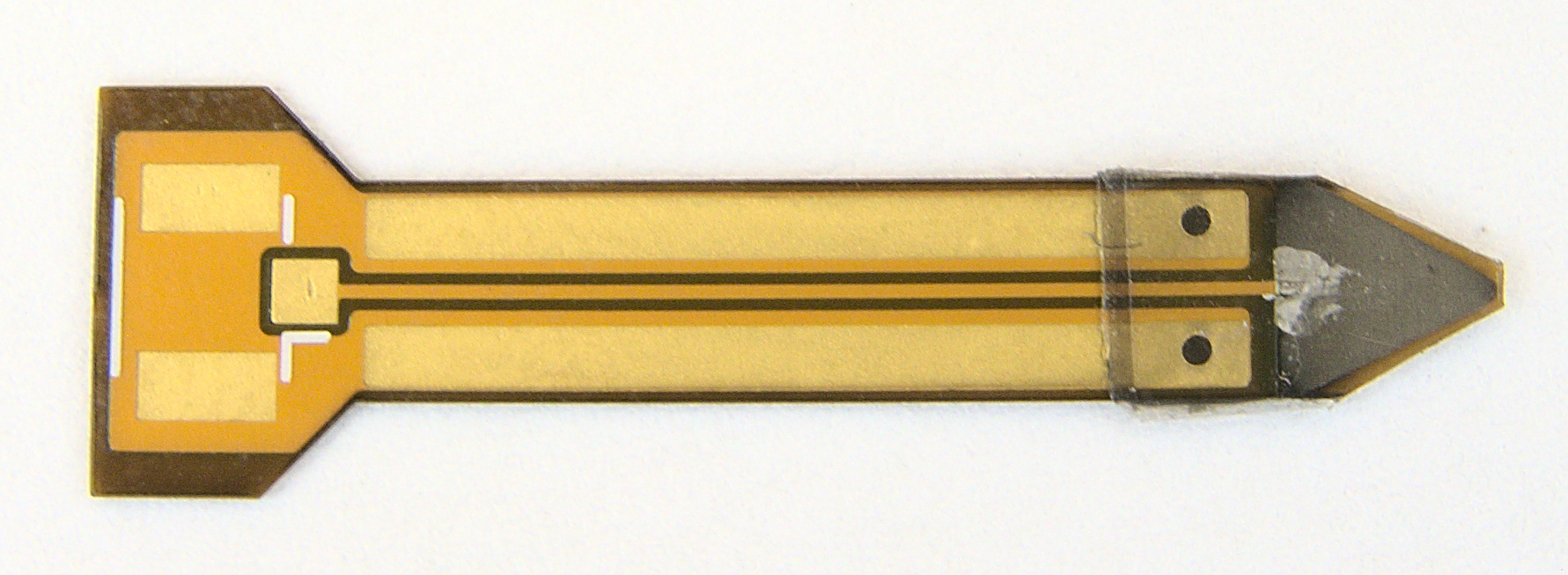}
    \caption{We trim the sensor to its final shape. Note the small border around the patterned electrode}
    \label{fig:fab_trim}
  \end{subfigure}

  \begin{subfigure}{\columnwidth}
    \includegraphics[width=\columnwidth]{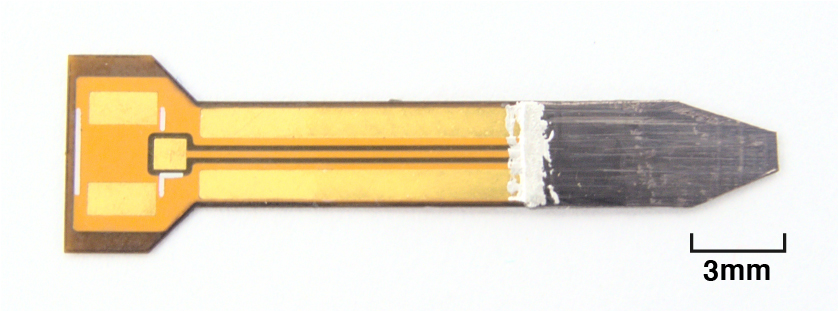}
    \caption{We sputter coat a second layer of \SI{200}{nm} aluminum onto the sensor, forming the ground shield.}
    \label{fig:fab_finished}
  \end{subfigure}

  \caption{The fabrication process for the UmboMic sensor.}
  \label{fig:fab}
\end{figure}

To simplify the fabrication of the UmboMic sensor, we use a flex PCB as the base substrate of the sensor. The custom flex PCB has a polyimide core with electrode and ground traces connecting to a U.FL connector solder footprint. We use photolithography to pattern triangular charge sense electrodes at the top of the flex PCB substrate, and this constitutes the active region of our sensor.

Through experimentation we found that cantilever designs with charge sense electrodes exposed to the outside of the sensor tend to have unacceptably high parasitic leakage conductance, especially in wet environments like the middle ear cavity.
Our fabrication strategy revolves around pre-patterning the charge sense electrodes and then trimming the sensor to leave a margin around the electrodes, eliminating this leakage path and improving the UmboMic apparatus noise floor.

We first apply \SI{200}{nm} of aluminum to both sides of the flex PCB using an AJA sputter coater. Next, we spin-coat a layer of AZ3312 positive photoresist on both sides of the sputter-coated PCB, bake for approximately two minutes at \SI{110}{\celsius}, place in a contact photolithography mask, and flood expose for 30 seconds on each side. We then dissolve the UV-exposed photoresist and the aluminum underneath using a tetramethyl ammonium hydroxide (TMAH) solution. Finally, we dissolve the remaining photoresist in acetone.
Figure~\ref{fig:fab_pattern} summarizes the stages of electrode deposition and patterning.

\subsection{PVDF adhesion}

Before gluing the PVDF film to the sputter coated metal, we reinforce the electrical connection between the patterned electrodes and the flex PCB traces with a silver conductive ink pen. We then sand one side of the PVDF with 3000-grit sandpaper to increase surface roughness and mask the portions of the flex PCB that must remain glue-free. Next, we generously apply epoxy between the two PVDF layers and the flex PCBs. Devcon Plastic Steel epoxy works well for bonding the PVDF to the polyimide substrate. We orient the piezoelectric films such that they have opposing polarization.
Finally, we squeeze as much epoxy as possible out from between the flex PCB and PVDF film with a doctor blade, and the stackup is left to cure.
This method achieves a \SI{10}{\um} epoxy thickness, which is sufficiently thin to allow efficient capacitive coupling.
%
%\marginpar{\tiny A side-view drawing would also be useful.}
%
The masking and bonding process is shown in Figure~\ref{fig:fab_tape}.

\subsection{Finishing steps}

After the epoxy is cured, we trim the PVDF and flex PCB to shape with scissors leaving a buffer of approximately \SI{300}{\um} between the edge of the electrode and the edge of the PCB layer of the sensor, shown in Figure~\ref{fig:fab_trim}.
This buffer serves to protect the electrodes from water ingress, which could otherwise short the sensor.
We then encapsulate the sensor tip with a \SI{200}{nm} layer of sputter-coated aluminum.
This outer layer serves as both a ground electrode and a ground shield, protecting the charge sense electrodes from EMI.
We connect the step between the PVDF and the flex PCB with conductive ink or adhesive as shown in Figure~\ref{fig:fab_finished}. This ensures the aluminum on the PVDF layer is electrically connected to the ground pad on the PCB at the tail. Finally, we solder a U.FL receptacle on either side of the tail end of the UmboMic sensor opposite from the electrodes. Figure~\ref{fig:diagram} shows the stack-up of the tip of the finished UmboMic sensor.

\begin{figure}[bt]
  \centering
  \includegraphics[width=\columnwidth]{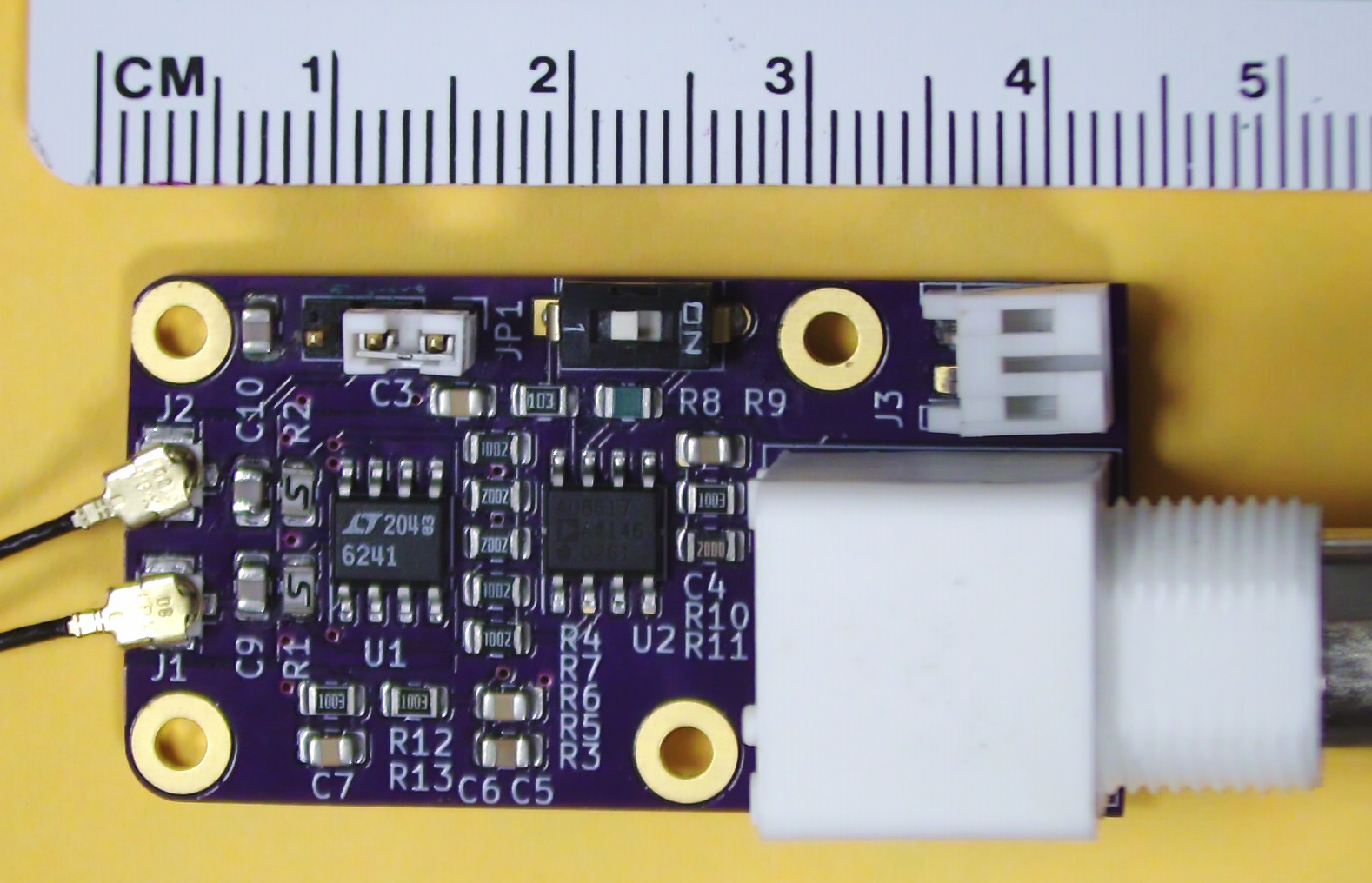}
  \caption{
    The assembled differential amplifier board.
  }
  \label{fig:diffamp_pic}
\end{figure}

\section{Differential Charge amplifier}
\label{sec:chargeamp}

  \begin{figure*}[ht]
  \centering
  \begin{circuitikz}[american, scale=1, font=\small]
    \draw[color=red, thick, dashed] (-6,2.2) 
        to[short, l={\small \hspace{0.97in} PVDF}] (-.55,2.2)
        -- (-.55,-2.2) -- (-6,-2.2) -- (-6,2.2);
  
    \draw(0,0.75)
    node[op amp, noinv input down, anchor=+](OA1){LTC6241};
    \draw(0,-0.75)
    node[op amp, noinv input up, anchor=+](OA2){LTC6241};
    \draw(0,0)
    (OA1.-) ++(-0.1,0) node[below]{$v_{\rm in+}$}
    (OA2.-) ++(-0.1,0) node[above]{$v_{\rm in-}$}

    (OA1.out) node[below]{$v_{\rm out-}$}
    (OA2.out) node[above]{$v_{\rm out+}$}
    (OA1.out) ++(-0.4,-0.4) node[below]{(oa1)}
    (OA2.out) ++(-0.4,+0.4) node[above]{(oa1)}

    (OA1.+) -- (OA2.+)
    (0,0) to[short, *-] ++(.15,0) node[ground, rotate=-90]{}
    (OA1.-) -- ++(0, 1.5) coordinate(C1)
    to[C, l_=$C_{\rm f}$, *-*] (C1 -| OA1.out) -- (OA1.out)
    (C1) -- ++(0, 1) coordinate(R1)
    to[R=$R_{\rm f}$] (R1 -| OA1.out) -- (C1 -| OA1.out)

    (OA2.-) -- ++(0, -1.5) coordinate(C2)
    to[C, l=$C_{\rm f}$, *-*] (C2 -| OA2.out) -- (OA2.out)
    (C2) -- ++(0, -1) coordinate(R2)
    to[R, l_=$R_{\rm f}$] (R2 -| OA2.out) -- (C2 -| OA2.out)
    ;

    \draw(OA2.out)
    ++(2, 0)
    node[op amp, noinv input down, anchor=+](OA3){AD8617}
    (OA1.out) to[R,a=\SI{10}{\kohm}, *-*] (OA1.out -| OA3.-) -- (OA3.-)
    (OA1.out -| OA3.-) to[R,a=\SI{10}{\kohm}] (OA1.out -| OA3.out) to[short, -*] (OA3.out)
     (OA2.out) to[R,a=\SI{10}{\kohm}, *-*] (OA2.out -| OA3.+) -- (OA3.+)
    to[R,a=\SI{10}{\kohm}] ++(0, -2.6) node[ground]{}

    (OA3.out) node[below]{$v_{\rm int}$}
    (OA3.out) ++(-0.4,-0.4) node[below]{(oa2)}
    ;

    \draw(OA3.out)
    (OA3.out) to[short,-] ++(1.5,0)
    node[op amp, noinv input up, anchor=+](OA4){AD8617}
    %(OA4.out) to[short, *-o] ++(0.5,0) node[above]{$v_{\rm out}$}
    (OA4.-) -- ++(0, -2) coordinate(R6)
    to[R,l^=$R_{\rm a}$, *-] (R6 -| OA4.out) -* (OA4.out)
    (R6) to[R,l_=$R_{\rm b}$, *-] ++(-2.4,0) coordinate(R7)
    to [C, l_=$C_{\rm b}$] ++(-0.7,0) coordinate(cb)
    to [short,-*] (cb -| OA3.-)
    (OA4.out) to[C,l^=$C_{\rm o}$, *-] ++(0,2.5)
    node[right]{$v_{\rm out}$}
    to[R, l^=$R_{\rm o}$, *-] ++(0,2.5)
    node[ground, rotate=180]{}
    (OA4.out) ++(-0.5,-0.4) node[below]{(oa3)}
    ;

    \draw(0,0)
    (-2,0) coordinate(CPZ)
    ++(-2,0) coordinate(CPR)
    ++(-1,0) coordinate(RPR)
    (OA2.-) to[short, i<=$i_{\rm in-}$, *-*] (OA2.- -| CPZ) -- (OA2.- -| RPR)
    to[R, l=$R_{\rm par}$] (OA1.- -| RPR) -- (OA1.- -| CPZ) to[short,i=$i_{\rm in+}$, *-*] (OA1.-)
    (OA2.- -| CPR) to[C, l_=$C_{\rm par}$, *-*] (OA1.- -| CPR)
    (OA2.- -| CPZ) to[voltage source, invert, l_=$v_{\rm piezo}$] (CPZ)
    to[short, i=$q_{\rm in}$] (CPZ)
    to[short, i_=$i_{\rm in}$] (CPZ)
    to[C, l_=$C_{\rm piezo}$] (OA1.- -| CPZ)

    (OA2.- -| CPR) to[short, -] ++(0,-0.4)
    to[C, l_=$C_{\rm gnd}$] ++(0, -1.7) node[ground]{}
    (OA1.- -| CPR) to[short, -] ++(0,0.4)
    to[C, l=$C_{\rm gnd}$] ++(0, 1.7) node[ground, rotate=180]{}
    ;
  \end{circuitikz}
  \caption{
    The differential sensor (outlined in red) and charge amplifier topology. We model the piezoelectric sensor as the voltage source $v_{\rm piezo}$ in series with the capacitor $C_{\rm piezo}$, together with a parasitic capacitor $C_{\rm par}$, leakage resistor $R_{\rm par}$, and capacitor to ground $C_{\rm gnd}$. Estimates of the piezoelecric and parasitic component values are given in Table \ref{parasitics}. Our implementation of the differential amplifier uses $R_{\rm f} = \SI{10}{G\ohm}$, $C_{\rm f} = \SI{1}{pF}$, $R_{\rm a} = \SI{90}{k\ohm}$, $R_{\rm b} = \SI{10}{k\ohm}$, $C_{\rm b} = \SI{100}{nF}$, $R_{\rm o} = \SI{100}{k\ohm}$, and $C_{\rm o} = \SI{100}{nF}$.
  }
  \label{fig:chargeamp}
\end{figure*}
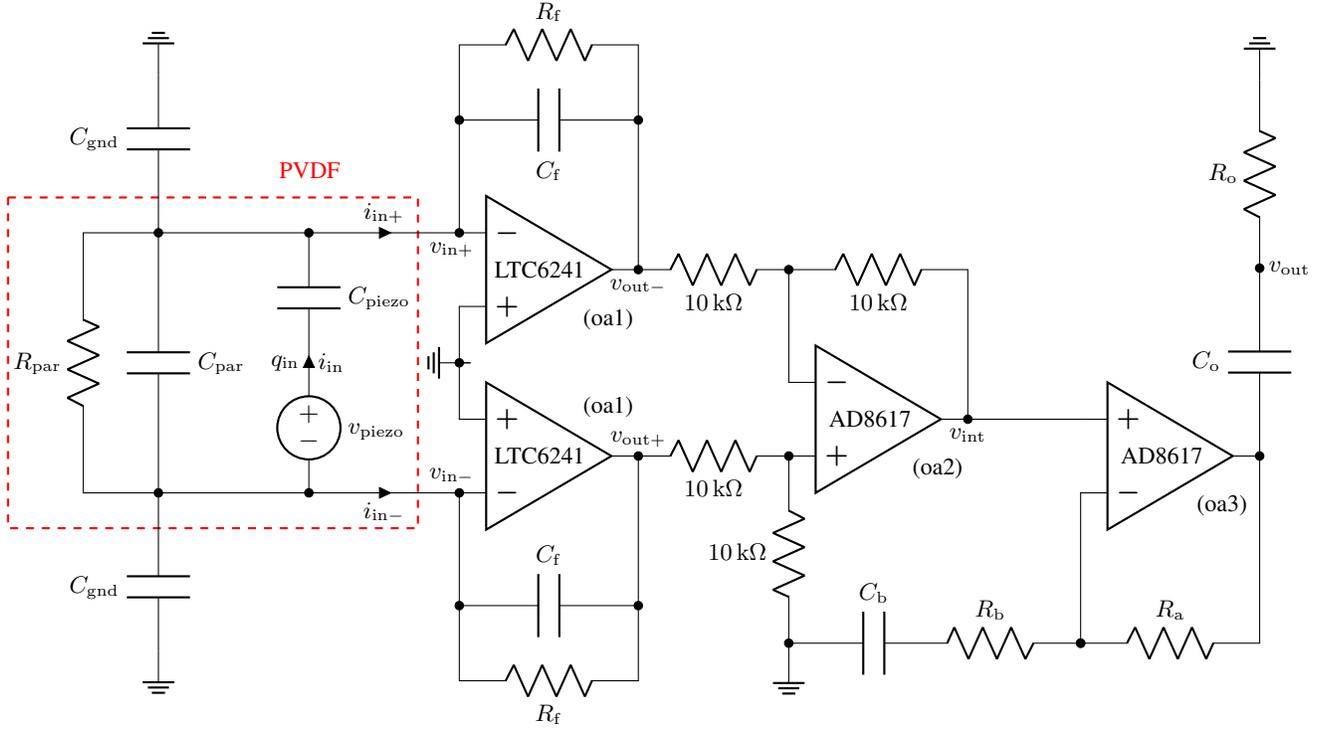

%\marginpar{\tiny vpiezo, Cpiezo, Cgnd, Rpar and Cpar should all be briefly explained in words.}
%
% A custom differential charge amplifier was developed in order to achieve a sufficiently low noise floor for a \SI{10}{pF} sensor.
It is imperative to device performance to achieve signal amplification without introducing too much noise. By developing our own differential charge amplifier, shown in Figures~\ref{fig:diffamp_pic} and \ref{fig:chargeamp}, we minimize the noise floor while providing a gain of \SI{20}{V/pC} over a \SI{-3}{dB} bandwidth of \SI{160}{Hz} to \SI{50}{kHz}.
Figure~\ref{fig:chargeamp} illustrates the charge amplifier connected to our differential sensor having capacitance $C_{\rm piezo}$ and charge output $q_{\rm in}$.
We also show parasitics $C_{\rm par}$, $R_{\rm par}$, and $C_{\rm gnd}$: parasitic parallel capacitor, leakage resistor, and capacitor to ground, respectively. Estimates of the piezoelectric and parasitic component values are given in Table~\ref{parasitics}. The charge-to-voltage gain of such a charge amplifier is invariant to parasitic resistance and capacitance, giving it good gain uniformity from sensor to sensor.
The amplifier's differential input interfaces with our differential-mode sensor to reduce EMI.
Similar differential charge amplifiers are frequently used as low-noise preamplifiers for high-impedance AC sources such as piezoelectric sensors~\cite{sirohi2000fundamental} and charged particle counters~\cite{hu_design_1998, duncan_electrostatic_2011}.
%We chose a differential charge amplifier as an analog front end for our microphone.
%We were unable to find commercially available charge amplifiers with sufficiently low noise floor for a \SI{10}{pF} sensor.
%Since we were already designing our own custom preamplifier, we used the differential charge amp topology shown in Figure~\ref{fig:chargeamp}.

\begin{table}[t]
    \centering
    \begin{tabular}{|c|c|c|c|}
    \hline $C_{\rm piezo}$ & $C_{\rm par}$ & $R_{\rm par}$ & $C_{\rm gnd}$ \\ \hline 10 pF & $\sim$1 pF & $\sim$ 1 T$\Omega$ & 0.6 fF \\ \hline 
    \end{tabular}
    \caption{Estimates of piezoelectric and parasitic component values.}
    \label{parasitics}
\end{table}

\subsection{Gain Analysis}

Our differential amplifier comprises two parallel low-impedance input stages based on the LTC6241 (dual LTC6240) op-amp (oa1), followed by a difference stage based on the AD8617 (dual 8613) op-amp (oa2), followed by a lead gain stage based on the AD8617 op-amp (oa3) with an output high-pass filter. The LTC6241 is chosen for its excellent noise performance. The AD8617 is chosen for its good noise performance, low bias current, and rail-to-rail operation. 

The amplifier input can be interpreted as either $q_{\rm in}$, $i_{\rm in}$ or $v_{\rm piezo}$, which are related by
\begin{equation}
    q_{\rm in} = i_{\rm in}/j\omega = C_{\rm piezo}v_{\rm piezo} \label{input} \quad .
\end{equation}
The internal high-pass charge-to-voltage gain $G_{\rm int}$ is then given by
\begin{equation}
v_{\rm int}/q_{\rm in} \equiv G_{\rm int} = \frac{2j\omega R_{\rm f}}{1 + j\omega R_{\rm f}C_{\rm f}} \label{gain}\quad ,
\end{equation}
which is independent of parasitics; note that the internal current-to-voltage gain is $G_{\rm int}/j\omega$. The overall gain of the amplifier is given by
\begin{eqnarray}
v_{\rm out}/q_{\rm in} & \equiv & G_{\rm out} \nonumber \\ & = & G_{\rm int} \frac{(1 + j\omega (R_{\rm a}+R_{\rm b})C_{\rm b})\ j\omega R_{\rm o}C_{\rm o}}{(1 + j\omega R_{\rm b}C_{\rm b})\ (1 + j\omega R_{\rm o}C_{\rm o})} \label{fullgain}\quad . \qquad
\end{eqnarray}
For the component values in Figure \ref{fig:chargeamp}, the highest high-pass cut-on frequency $1/R_{\rm b}C_{\rm b}$ is set at 1000 rad/s (160~Hz) to filter out low-frequency body noise. The high-end low-pass cut-off frequency is set well above the audio range by the op-amp dynamics, and so is not modeled here. Finally, the mid-band gain is given by $G_{\rm out} = 20/C_{f} = 20$ V/pC.

\subsection{Noise analysis}

There are five significant noise sources in the amplifier: Johnson noise from $R_{\rm f}$ and $R_{\rm par}$, voltage noise and current noise from oa1, and voltage noise from oa2. Johnson noise is treated here as a parallel current source. Being in parallel with the input current, $R_{\rm par}$ contributes an input-referred current variance density of $4k_{\rm B}T/R_{\rm par}$. Together, the two $R_{\rm f}$ contribute the same input-referred current variance density as would a single $2R_{\rm f}$ in parallel with the input current, or $4k_{\rm B}T/2R_{\rm f}$. Thus, the total input-referred current variance density associated with Johnson noise is
\begin{equation}
  \overline{i^2}_{\rm in,johnson} = \frac{4 k_{\rm B} T}{(2 R_{\rm f}) \parallel R_{\rm par}} \quad .
  \label{johnson_noise}
\end{equation}

The noise voltages at $v_{\rm in+}$ and $v_{\rm in-}$ effectively produce an input-referred  noise current $i_{\rm in,v,oa1}$ determined by the impedances of the sensor and the oa1 feedback network. While the noise voltages are not completely frequency independent, flicker noise for the LTC6240 is negligible above \SI{100}{Hz}. Thus, the noise voltages are modeled here as white noise sources. Recognizing that the overall amplifier will reject common-mode voltage noise, define
\begin{equation}
    v_{\rm diff,v,oa1} = v_{\rm in-,v,oa1} -\ v_{\rm in+,v,oa1} \quad , \label{vdiff}
\end{equation}
\begin{equation}
    Z_{\rm diff} = R_{\rm par} \parallel \frac{1}{j \omega (C_{\rm piezo} + C_{\rm par} + C_{\rm gnd}/2)} \label{Zdiff}
\end{equation}
and
\begin{equation}
   Z_{\rm f} = R_{\rm f} \parallel \frac{1}{j\omega C_{\rm f}} \quad , \label{Zf} 
\end{equation}
where $v_{\rm in+,v,oa1}$ and $v_{\rm in-,v,oa1}$ are the noise voltages at $v_{\rm in+}$ and $v_{\rm in-}$, respectively. Then, $v_{\rm diff,v,oa1}$ drives the internal voltage
\begin{equation}
    v_{\rm int,v,oa1} = \frac{Z_{\rm diff}+2 Z_{\rm f}}{Z_{\rm diff}}\ v_{\rm diff,v,oa1} \quad ,
\end{equation}
with corresponding input-referred noise current
\begin{equation}
    i_{\rm in,v,oa1} = \frac{j\omega}{G_{\rm int}} \cdot \frac{Z_{\rm diff}+2 Z_{\rm f}}{Z_{\rm diff}}\ v_{\rm diff,v,oa1} \quad . 
    \label{v_oa1_noise}
\end{equation}
Finally substitution of (\ref{Zdiff}) and (\ref{Zf}) into (\ref{v_oa1_noise}), and recognition that the two op-amp noise voltages are independent, yields
\begin{equation}
\overline{i^2}_{\rm in,v,oa1} = \omega^2\ \left| \frac{Z_{\rm diff}+2Z_{\rm f}}{G_{\rm int}\ Z_{\rm diff}}\right|^2\ 2\ \overline{v^2}_{\rm oa1} \quad , \label{v_oa1_noise_density}
\end{equation}
as the corresponding input-referred current variance density, where $\overline{v^2}_{\rm oa1}$ is the voltage variance density of oa1.

%Defining
%\begin{equation}
%  Z_{\rm comm} = R_{\rm f} \parallel \frac{1}{j \omega (C_{\rm f} + C_{\rm gnd})}
%  \label{eq:zcomm}
%\end{equation}
%and
%\begin{equation}
%  Z_{\rm diff} = R_{\rm par} \parallel \frac{1}{j \omega (C_{\rm par} + C_{\rm piezo})}\quad,
%  \label{eq:zdiff}
%\end{equation}
%the induced noise currents are
%\begin{equation}
%  i_{\rm in,op1,v+} = \frac{v_{\rm op1,-} - v_{\rm op1,+}}{Z_{\rm diff}} - \frac{v_{\rm op1,+}}{Z_{\rm comm}},
%\end{equation}
%with $i_{\rm in,op1,v-}$ defined equivalently. Next, define the differential input current $i_{\rm in} = \frac{1}{2} (i_{\rm in,+} - i_{\rm in,-})$ so that
%\begin{equation}
%  i_{\rm in,op1,v} = (v_{\rm op1,-} - v_{\rm op1,+}) \cdot \left(\frac{1}{Z_{\rm diff}} + \frac{1}{2Z_{\rm comm}} \right)\quad .
%\end{equation}
%Finally, because the two op amps are independent noise sources, the differential induced noise current has magnitude
%\begin{equation}
%  \overline{i^2}_{\rm in,op1,v} = 2 \overline{v^2}_{\rm op1} \cdot \left| \frac{1}{2 Z_{\rm comm}} + \frac{1}{Z_{\rm diff}} \right|^2\quad .
%  \label{eq:voltage_noise}
%\end{equation}

Each noise current of the LTC6240 can be modeled using
\begin{equation}
\overline{i^2}_{\rm oa1} = \overline{\iota^2}_{\rm oa1} + \omega^2 \overline{q^2}_{\rm oa1}
\end{equation}
where $\iota_{\rm oa1}$ and $q_{\rm oa1}$ are both constants \cite{ltc6240}. It is further assumed that $v_{\rm oa1}$, $\iota_{\rm oa1}$ and $q_{\rm oa1}$ are all independent; the correlation between op amp current noise and voltage noise is unspecified in~\cite{ltc6240}. When two op amps are used to construct a differential amplifier that rejects common-mode current noise, the resulting input-referred current variance density becomes
\begin{equation}
  \overline{i^2}_{\rm in,i,oa1} = \frac{\overline{\iota^2}_{\rm oa1} + \omega^2 \overline{q^2}_{\rm oa1}}{2} \quad .
  \label{current_noise}
\end{equation}

Finally, the input-referred current variance density resulting from the difference stage may be expressed as
\begin{equation}
  \overline{i^2}_{\rm in,v,oa2} = \frac{4 \omega^2 \overline{v^2}_{\rm oa2}}{|G_{\rm int}|^2} = \overline{v^2}_{\rm oa2} \left(\frac{1}{R_{\rm f}^2} + \omega^2 C_{\rm f}^2 \right) \quad .
  \label{stage2_noise}
\end{equation}

The total input-referred current variance density $\overline{i^2}_{\rm in}$ is obtained by summing (\ref{johnson_noise}), (\ref{v_oa1_noise_density}), (\ref{current_noise}) and (\ref{stage2_noise}). Dividing $\overline{i^2}_{\rm in}$ by $\omega^2$ gives the input-referred charge variance density $\overline{q^2}_{\rm in}$. Finally, expanding $Z_{\rm diff}$ and $Z_{\rm f}$, and collecting terms, gives 
\begin{multline}
  \overline{q^2}_{\rm in} = 2 \overline{v^2}_{\rm oa1} C_{\rm tot}^2 + \frac{1}{2} \overline{q^2}_{\rm oa1} + 
  \overline{v^2}_{\rm oa2} C_{\rm f}^2 \\
  + \frac{1}{\omega^2} \left(
    \frac{4 k_B T}{(2 R_{\rm f}) \parallel R_{\rm par}} + \frac{\overline{\iota^2}_{\rm oa1}}{2} + \right. \\
  \left. + \frac{2 \overline{v^2}_{\rm oa1}}{((2 R_{\rm f}) \parallel R_{\rm par})^2} +
    \frac{\overline{v^2}_{\rm oa2}}{R_{\rm f}^2}
  \right)\quad , \label{allnoise}
  \end{multline}
  where
  \begin{equation}
    C_{\rm tot} = C_{\rm piezo} + C_{\rm par} + \frac{1}{2} C_{\rm f} + \frac{1}{2} C_{\rm gnd} \quad.
  \label{eq:input_noise}
  \end{equation}
  From this point forward, $\sqrt{\overline{q^2}_{\rm in}}$ is referred to as the equivalent noise charge (ENC) density.

\subsection{Practical component selection}

Important design guidelines can be extracted from (\ref{allnoise}) and (\ref{eq:input_noise}). Parasitic leakage conductance and capacitance are universally bad from a noise perspective, and should be minimized for any given sensor design. Minimizing parasitic capacitance is especially important, as the $2 \overline{v^2_{\rm oa1}} C^2_{\rm tot}$ term in  (\ref{allnoise}) is a significant part of the amplifier noise floor. Furthermore, the ratio of $C_{\rm piezo}$ to $C_{\rm f}$ is effectively the voltage gain of the first stage; $C_{\rm f}$ should be several times smaller than $C_{\rm piezo}$ to minimize the second-stage contribution to the noise floor. We have built working prototypes with $C_{\rm f}$ up to \SI{10}{pF}, but those with $C_{\rm f} = \SI{1}{pF}$ work quite well. Since the differential charge amplifier requires good matching between the two input stages to achieve an acceptable common-mode rejection ratio, we use PCB capacitors to implement $C_{\rm f}$. By using a four-layer PCB and building the capacitors between the bottom two layers, we can implement each capacitor in a $3$~mm $\times$ 3 mm area with good matching and shielding.

The value of $R_{\rm f}$ requires more care. Ideally, $R_{\rm f}$ should be as large as possible but increased $R_{\rm f}$ gives worse bias stability. We observed that increasing $R_{\rm f}$ beyond 10~G$\Omega$ does not yield significant performance benefits.

The centerpiece of the amplifier is the low-noise op amp used for the first stage, as this sets the absolute lower bound on the noise floor. Choosing this op amp based on (\ref{allnoise}) requires balancing $\overline{v_{\rm in}^2}$ and $\overline{i_{\rm in}^2}$ over the desired frequency range and sensor capacitance. This requirement rules out op amps with bipolar or JFET input stages because these op amps typically have unacceptably high current noise. Op amps with CMOS input stages have voltage noise several times higher than top-of-the-line JFET or bipolar op amps, but with far lower current noise. Of these, the LTC6240 appears to offer the best combination of voltage noise and current noise, with the LTC6081 and LTC6078 providing respectable performance with lower power consumption. Previous use of the LT1792, which has significantly worse current noise than the LTC6240, caused the current noise to dominate  the sensor noise floor at low frequencies. See Table \ref{opamps} for an op amp comparison.

The second-stage difference amplifier requirements are far more relaxed. The AD8617 has a noise floor of approximately \SI{25}{nV/\sqrt{Hz}}, and so contributes \SI{50}{nV/\sqrt{Hz}} to $v_{\rm int}$. Each \SI{10}{\kohm} resistor contributes \SI{13}{nV/\sqrt{Hz}}. The total noise contribution is therefore \SI{56}{nV/\sqrt{Hz}}. Using (\ref{gain}) gives an input-referred white noise contribution of only \SI{0.028}{aC/\sqrt{Hz}}, which is insignificant compared to the noise floor of the complete amplifier.

\begin{table}
  \small
  \centering
  \begin{tabular}{c | c | c | c | c}
    \textbf{Property} & \textbf{LT1792} & \textbf{LTC6240} & \textbf{LTC6081} & \textbf{LTC6078} \\ \hline
    $\bar v_{\rm in}$ (\si{nV/\sqrt{Hz}}) & 4.2 & 7    & 13  & 18 \\
    $\bar i_{\rm in}$ (\si{fA/\sqrt{Hz}}) & 10  & 0.56 & 0.5 & 0.56 \\
    $C_{\rm in}$ & \SI{27}{pF} & \SI{3.5}{pF} & \SI{3}{pF} & \SI{10}{pF} \\
    Power & \SI{76}{mW} & \SI{6.7}{mW} & \SI{1.2}{mW} & \SI{200}{\micro W} \\
  \end{tabular}
  \vspace{0.1in}
  \caption{
    For small capacitance devices, the LTC6240 offers the best performance, at the expense of power consumption compared to the LTC6081 and LTC6078.
  }
  \label{opamps}
\end{table}

% Equation~\ref{eq:noise floor} shows that the noise floor of the device is fundamentally limited by the piezoelectric dielectric loss and the noise performance of the first-stage op-amp.

% To optimize noise performance, the amplifier is built around the LTC6241 low-noise op-amp \cite{ltc6240}.
% This op-amp has very low noise voltage and noise current, allowing for a rough noise impedance match with the sensor.
% Additionally, $C_{\rm f} = \SI{1}{pF}$ is chosen to reduce parasitic capacitance, and $R_{\rm f} = \SI{10}{G\ohm}$ is chosen to reduce resistor-induced thermal noise.
% Each $C_{\rm f}$ is implemented with a $\SI{2}{mm} \times \SI{2}{mm}$ PCB interconnect on a four-layer board. The amplifier was quite stable and reliable in tandem with the waterproof cantilever design.

\begin{figure}
  \centering
  \includegraphics[width=\columnwidth]{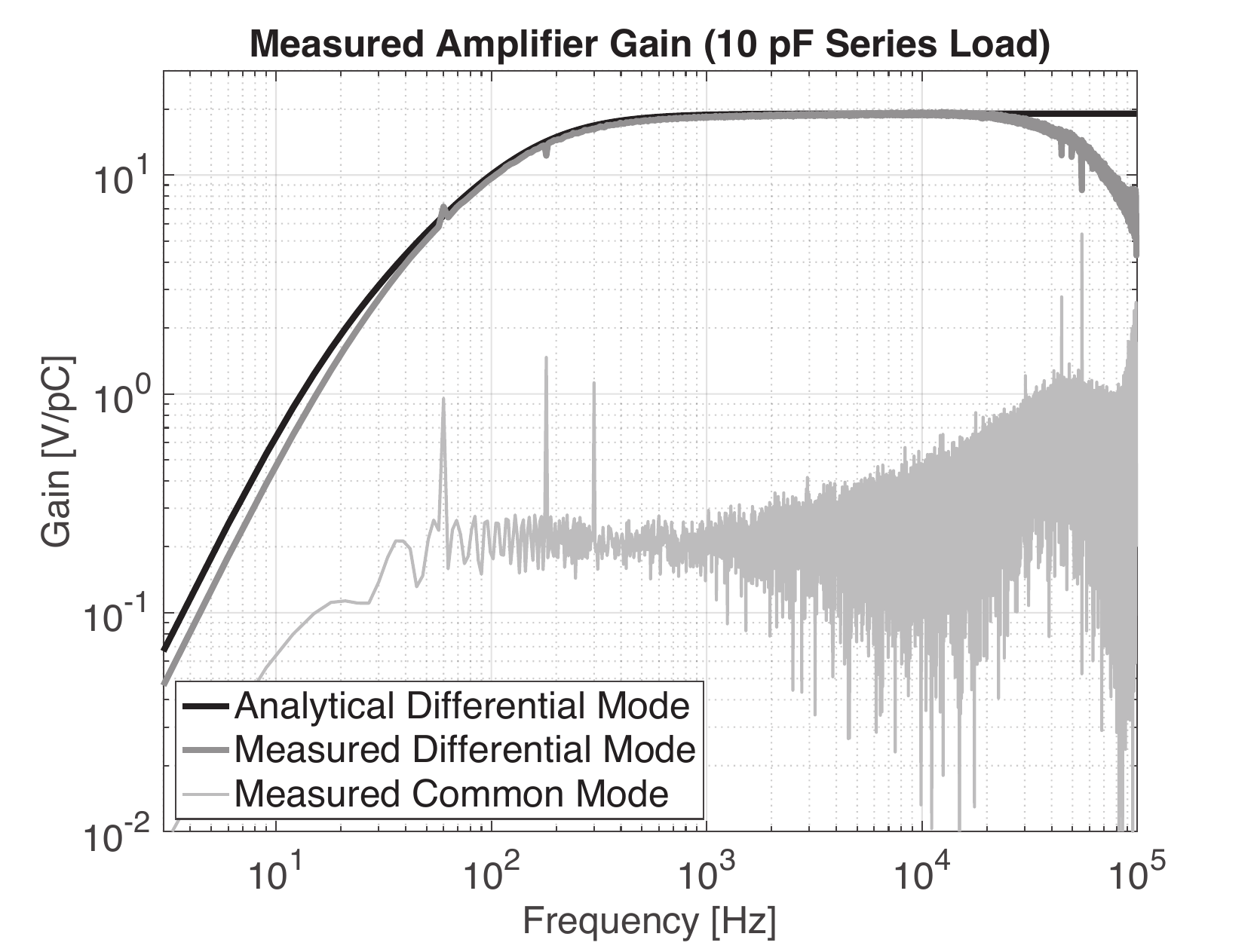}
    \caption{We measure gain by connecting a \SI{10}{pF} capacitor in series with each input. We achieved a charge gain of \SI{1.91e13}{V / C}, within \SI{5}{\percent} of the design gain.}
    \label{fig:diffamp_gain}
  \end{figure}

  \begin{figure}
    \includegraphics[width=\columnwidth]{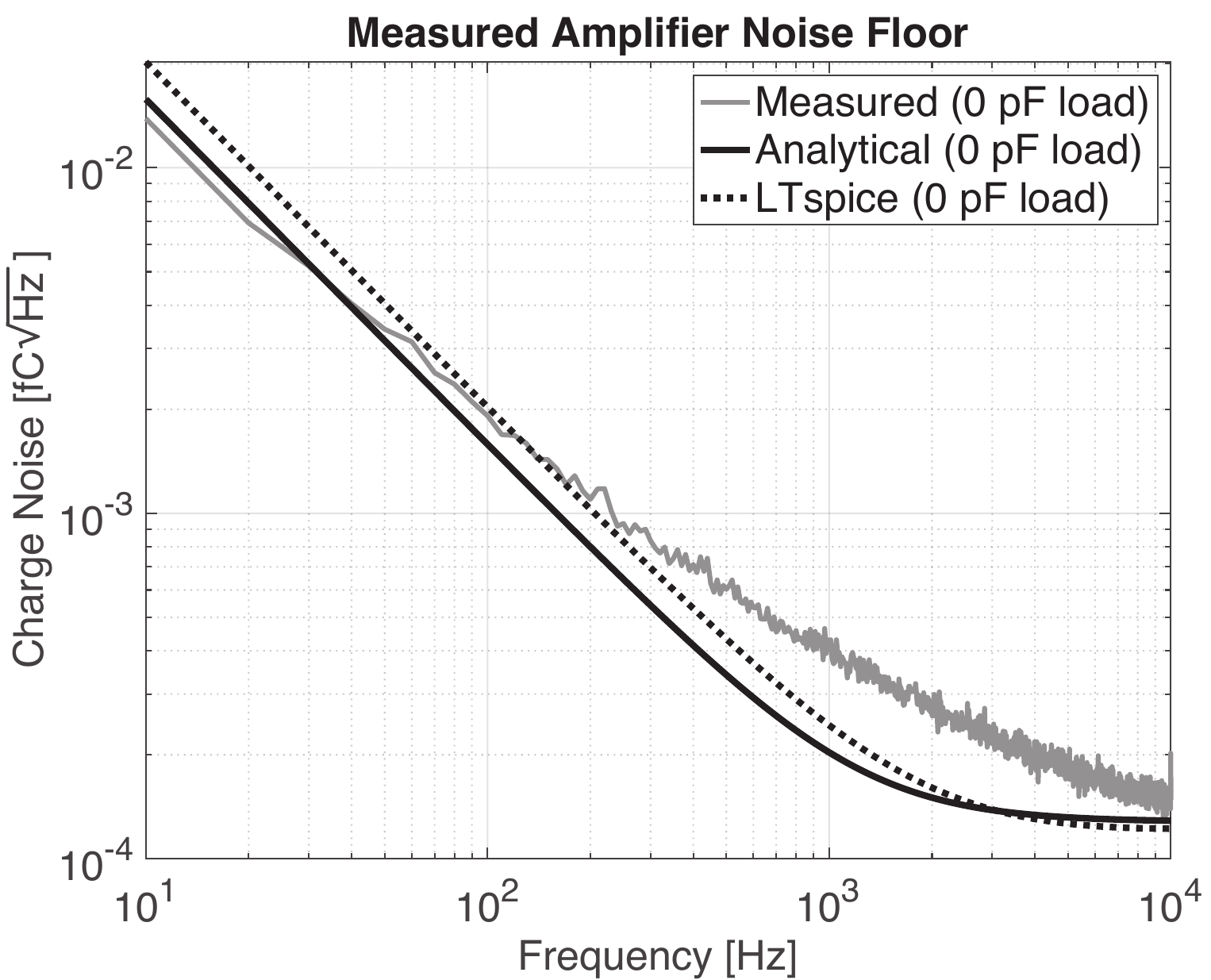}
    \caption{The output voltage noise of the amplifier under no load on its high gain setting (\SI{20}{V / pC}).
    The amplifier achieves an equivalent noise charge (ENC) of roughly \SI{3}{fC} over a bandwidth of \SI{200}{Hz}~to~\SI{20}{kHz}.}
    \label{fig:diffamp_noise}
  \end{figure}

  \begin{figure}
    \includegraphics[width=\columnwidth]{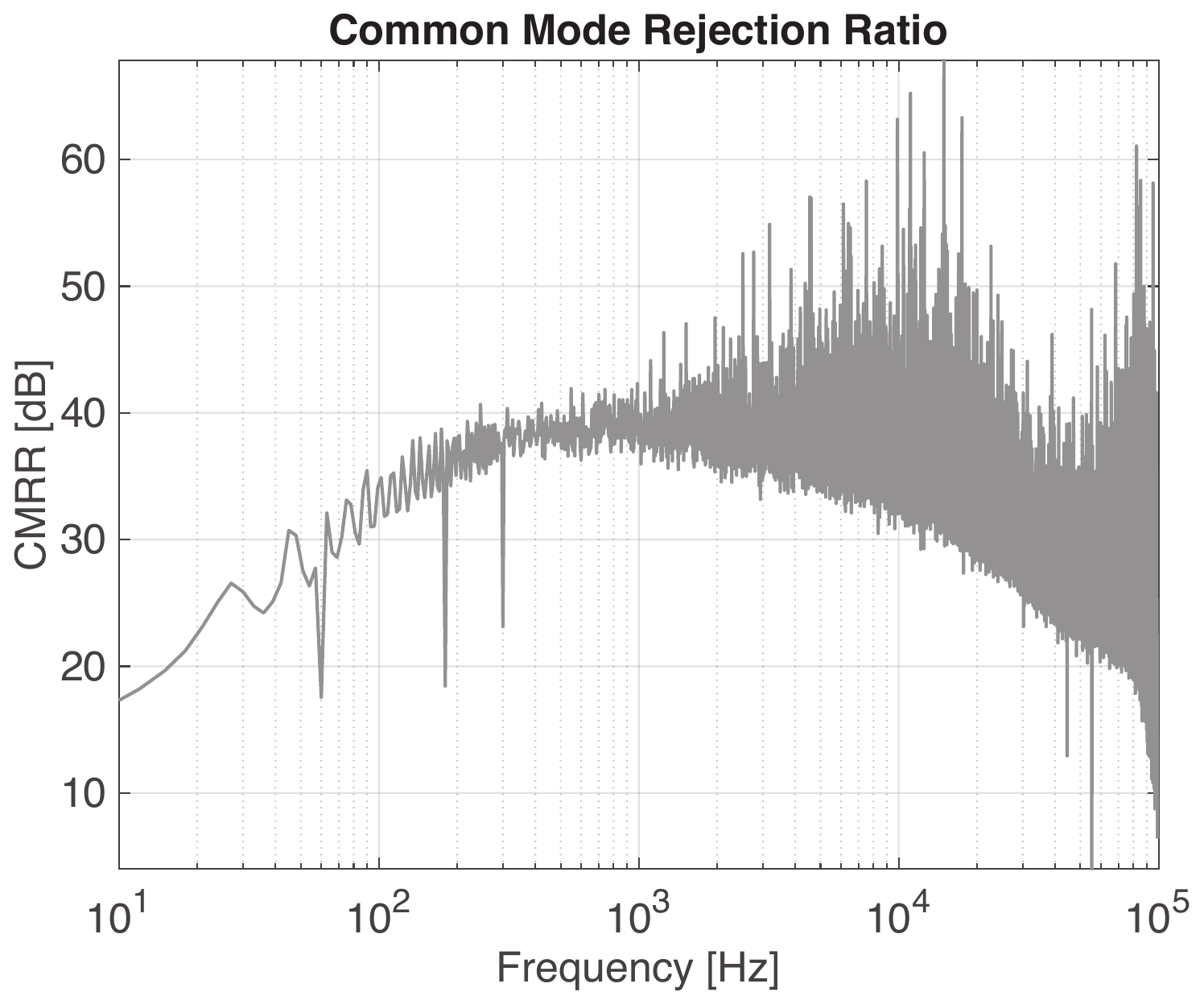}
    \caption{A common-mode rejection ratio of 30~to~\SI{40}{dB} is a reasonable expectation for this amplifier. The spikes at 60, 180, and 300~\si{Hz} appear to be due to poor shielding and long cable runs to test equipment.}
    \label{fig:cmrr}
  \end{figure}

\subsection{Specifications}

Our amplifier has a measured gain of \SI{19.1}{V/pC} over a \SI{-3}{dB} bandwidth of \SI{160}{Hz} to \SI{50}{kHz}.
This comes to within \SI{5}{\percent} of our \SI{20}{V/pC} target gain and exceeds our minimum target bandwidth of \SI{200}{Hz} to \SI{20}{kHz}.
We measured an equivalent noise charge over this target bandwidth of \SI{30}{aC} (\SI{185}{e^-}) with no sensor attached.
With one of our sensors attached, we measured the noise floor to be \SI{62}{aC} (\SI{385}{e^-}). Figure \ref{fig:diffamp_gain} shows the transfer function of our amplifier; Figure \ref{fig:diffamp_noise} shows its noise floor while unloaded and loaded with our sensor. Note that the analytically derived noise floor closely matches the noise floor simulated using LTspice.

The principal reason for building a custom charge amplifier is the lack of commercial low-noise amplifiers available for low-capacitance sensors.
Table~\ref{tab:comparison} illustrates this by comparison.
The CEC 1-328 is the highest-performing commercially available differential charge amp we could find, while the Femto HQA-15M-10T is the best available single-ended charge amp.
Our amplifier outperformed both, although their datasheets did not clearly specify the test load capacitance or spectral noise density.
We also found references to charge amplifiers in the literature.
The singled-ended charge amp inside the ELectrostatic Dust Analyzer (ELDA)~\cite{duncan_electrostatic_2011,xie_laboratory_2013} used an LTC6240 and performed similarly to our design, while Kelz~et.~al.~\cite{kelz_fully_2018} created a fully-integrated differential charge amp with excellent noise performance.

\begin{table}
  \centering
  \begin{tabular}{c | c | c | c }
    \textbf{Amplifier} & \textbf{q\textsubscript{noise}} & \textbf{C\textsubscript{load}} & \textbf{Bandwidth} \\ \hline
    Our amplifier & \SI{30}{aC} & \SI{0}{pF} & \SI{160}{Hz}--\SI{20}{kHz} \\ \hline
    Our amplifier & \SI{62}{aC} & \SI{25}{pF} & \SI{160}{Hz}--\SI{20}{kHz}\\ \hline
    CEC 1-328~\cite{cec_amp} & \SI{500}{aC} & N/A & \SI{5}{Hz}--\SI{10}{kHz} \\ \hline
    Femto HQA-15M-10T~\cite{femto_amp} & \SI{350}{aC} & N/A & \SI{250}{Hz}--\SI{15}{MHz} \\ \hline
    ELDA (LTC6240)~\cite{duncan_electrostatic_2011,xie_laboratory_2013} & \SI{57}{aC} & \SI{5}{pF} & \SI{7}{Hz}--\SI{10}{kHz} \\ \hline
    Kelz et. al.~\cite{kelz_fully_2018} & \SI{18}{aC} & \SI{5.4}{pF} & \SI{7}{Hz}--\SI{300}{kHz}
  \end{tabular}
  \caption{A comparison of this paper's preamp to charge amplifiers available commercially and in literature.  Some datasheets did not specify load capacitance.  Our amplifier demonstrates good input--referred noise charge at audio frequencies.}
  \label{tab:comparison}
\end{table}
  
\section{Measurement techniques}

We test our sensors in fresh-frozen cadaveric human temporal bones (no chemical preservatives) and conduct all measurements inside a soundproof and electrically isolated room at the Massachusetts Eye and Ear (MEE).
This allows us to take accurate measurements without background electrical, vibrational, or acoustic noise.
Fresh cadaveric human temporal bones are procured through Massachusetts General Hospital.
% Figure~\ref{fig:bench} shows the bench-testing setup, where the sensor is actuated with a smoothed tip glass rod glued to a piezoelectric stack.
% The motion of the piezoelectric stack was measured by laser Doppler vibrometry (LDV) to be \SI{16.9}{nm/V}. {\color{red} Include bench testing plot or take out mention of bench testing.}

% \begin{figure}[ht]
%   \centering
%   \includegraphics[width=\columnwidth]{images/v2.0_piezo_stack_labeled.png}
%   \caption{Bench testing: The sensor is held in a 3D-printed clamp and stimulated with a glass rod glued to a piezo stack.}
%   \label{fig:bench}
% \end{figure}

\begin{figure}
  \centering
  \includegraphics[width=\columnwidth]{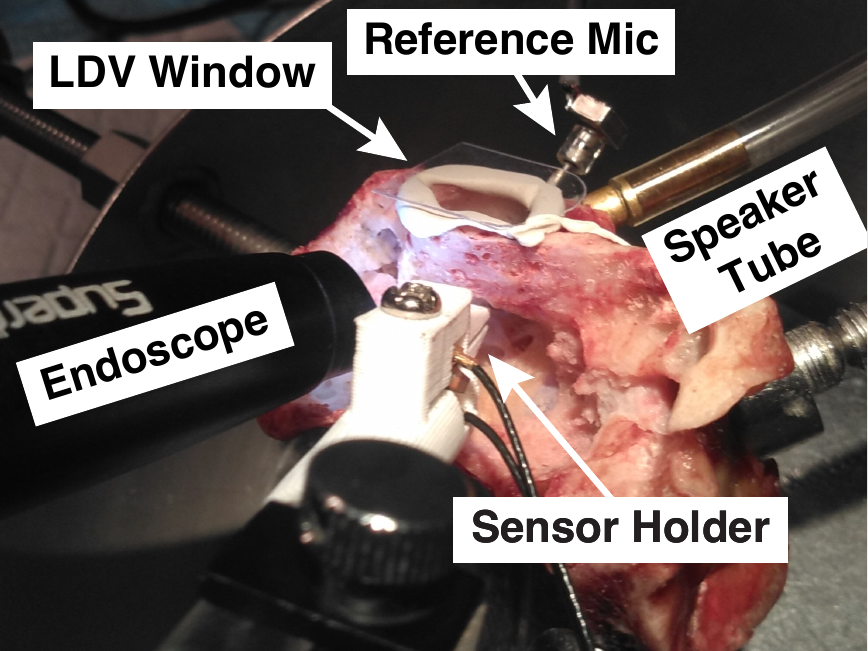}
  \caption{
    A 3D-printed clamp holds the sensor under the umbo.
    A speaker introduces sound pressure to the sealed ear canal.
  }
  \label{fig:tb_testing}
\end{figure}

Figure~\ref{fig:tb_testing} shows our temporal bone test setup.
A 3D-printed clamp holds the UmboMic sensor under the umbo while a transparent film of plastic seals the ear canal.
An external speaker introduces a sound pressure stimulus to the ear canal -- typically a sinusoidal sweep from \SI{100}{Hz} to \SI{20}{kHz}.
A calibrated Knowles EK3103 probe-tube reference microphone measures this sound pressure stimulus, with the probe tube opening directly above the eardrum. We measure over a range of ear canal pressure, from approximately \SI{60}{dB} to \SI{100}{dB} SPL (in the linear range). Umbo velocity at the tympanic membrane is measured with a laser Doppler vibrometry (LDV) beam through a clear window covering the ear canal.

The noise floor is measured by taking a Fourier transform of several seconds of amplifier noise with the sensor attached. Then, the Fourier transform is smoothed and normalized to a 1/3-octave bandwidth to permit direct comparison of the noise floor and sensitivity in the same graph, as shown in Figure~\ref{fig:freqresp}.

We also measure EMI sensitivity by placing the UmboMic sensor inside an aluminum foil ball without the sensor touching the foil. The foil is connected to a voltage source, thus placing the UmboMic sensor inside a nearly uniform electric potential.
Because our charge amplifier has a well-defined charge-to-voltage gain, we can accurately compute the ``EMI capacitance" of the UmboMic sensor, namely the charge induced by an external potential, and hence $C_{\rm gnd}$.

\section{Results}

\begin{figure}
  \centering
  \includegraphics[width=\columnwidth]{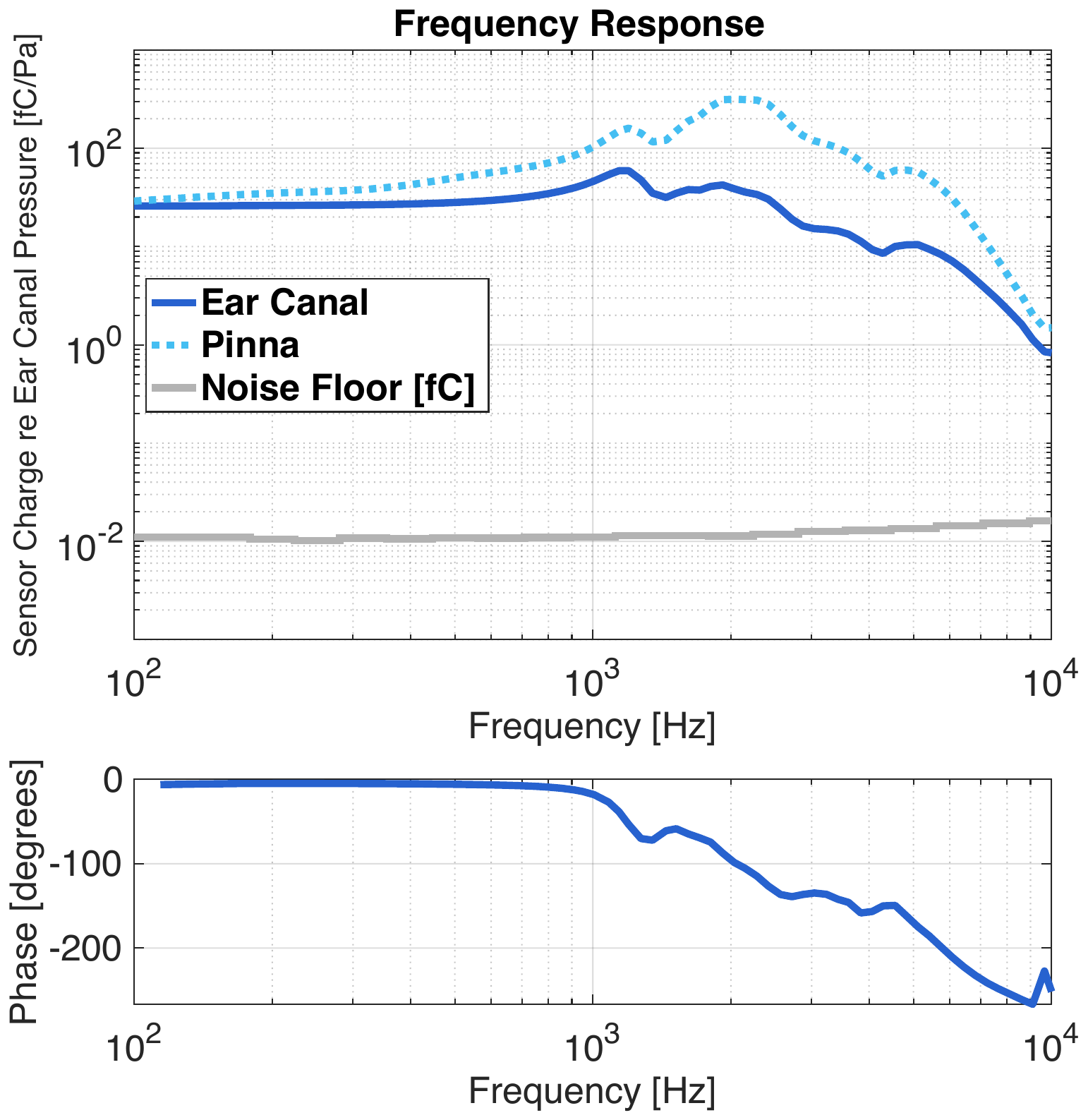}
  \caption{
    The frequency response of the UmboMic appartus relative to ear canal pressure.
    Extrapolated response including gain from the pinna (\SI{45}{\degree} azimuth) is also shown. Note that the noise floor is in units of \SI{}{fC} and not normalized by ear canal pressure.
  }
  \label{fig:freqresp}
\end{figure}

\begin{figure}
  \centering
  \includegraphics[width=\columnwidth]{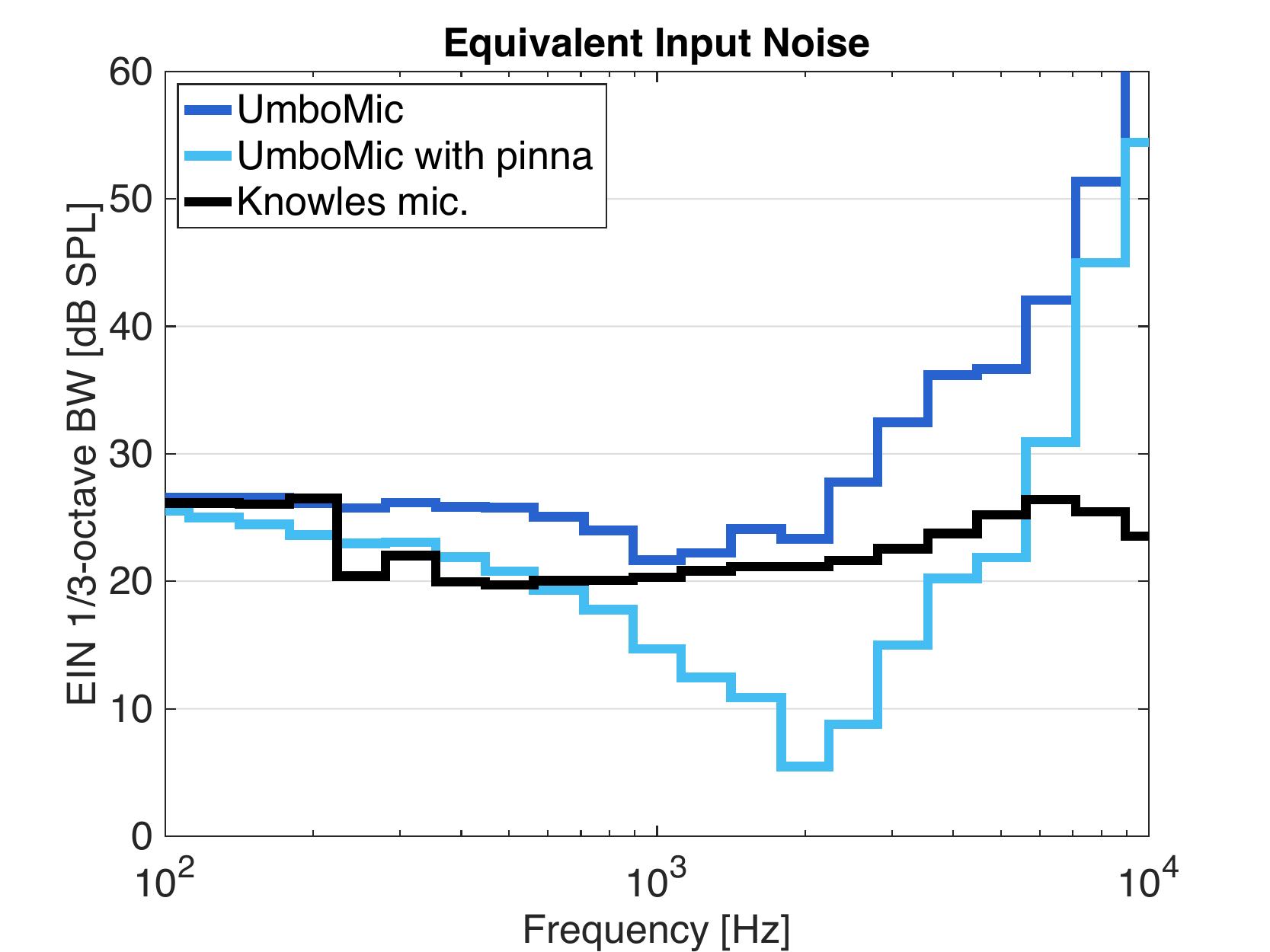}
  \caption{Equivalent input noise (EIN) normalized to 1/3-octave bandwidth.}
  \label{fig:ein}
\end{figure}

\begin{figure}[ht]
  \centering
  \includegraphics[width=\columnwidth]{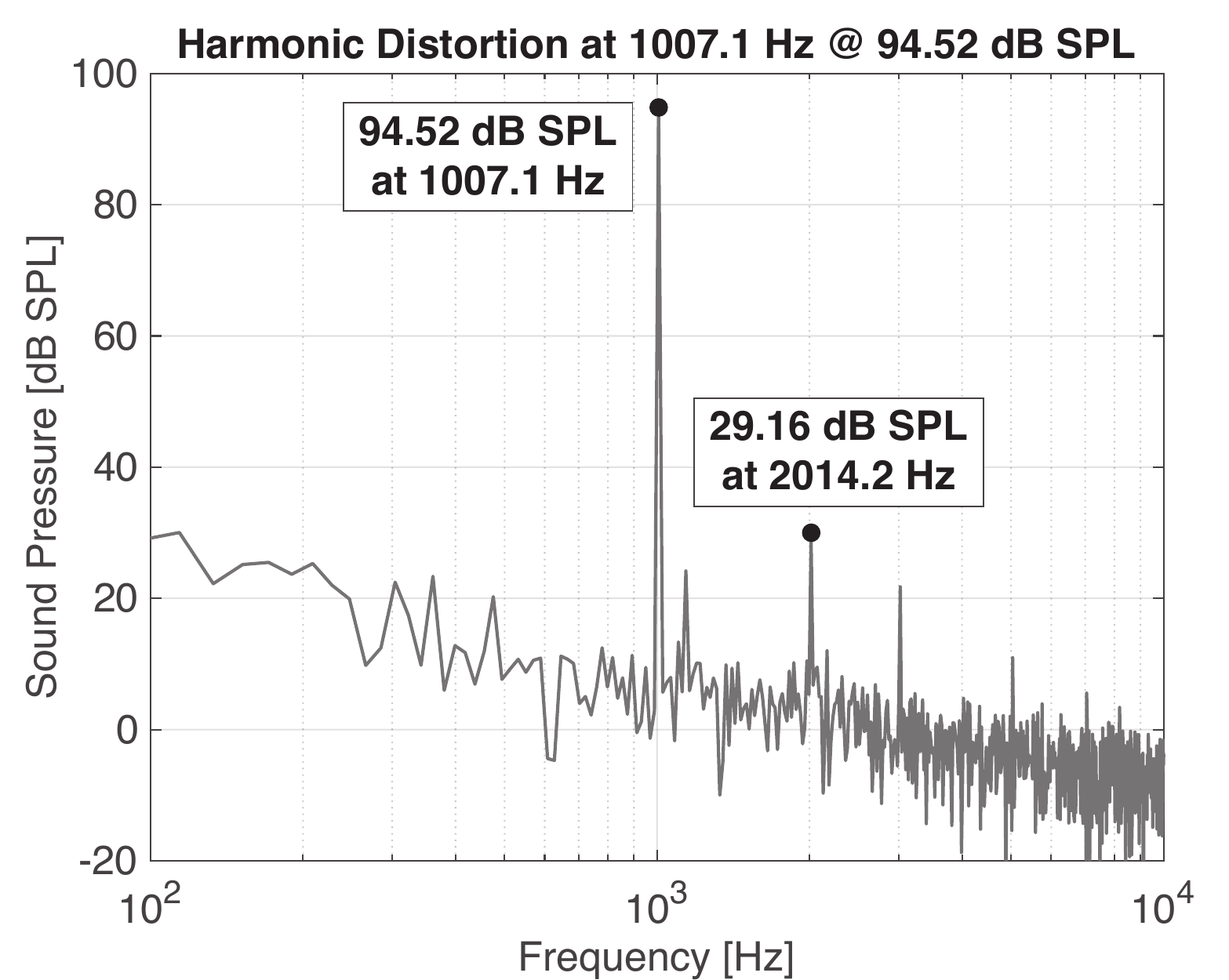}
  \caption{Harmonic distortion of less than \SI{0.1}{\percent} is demonstrated at \SI{94.5}{dB} SPL.}
  \label{fig:harmdist}
\end{figure}

\begin{figure}[ht]
  \centering
  \includegraphics[width=\columnwidth]{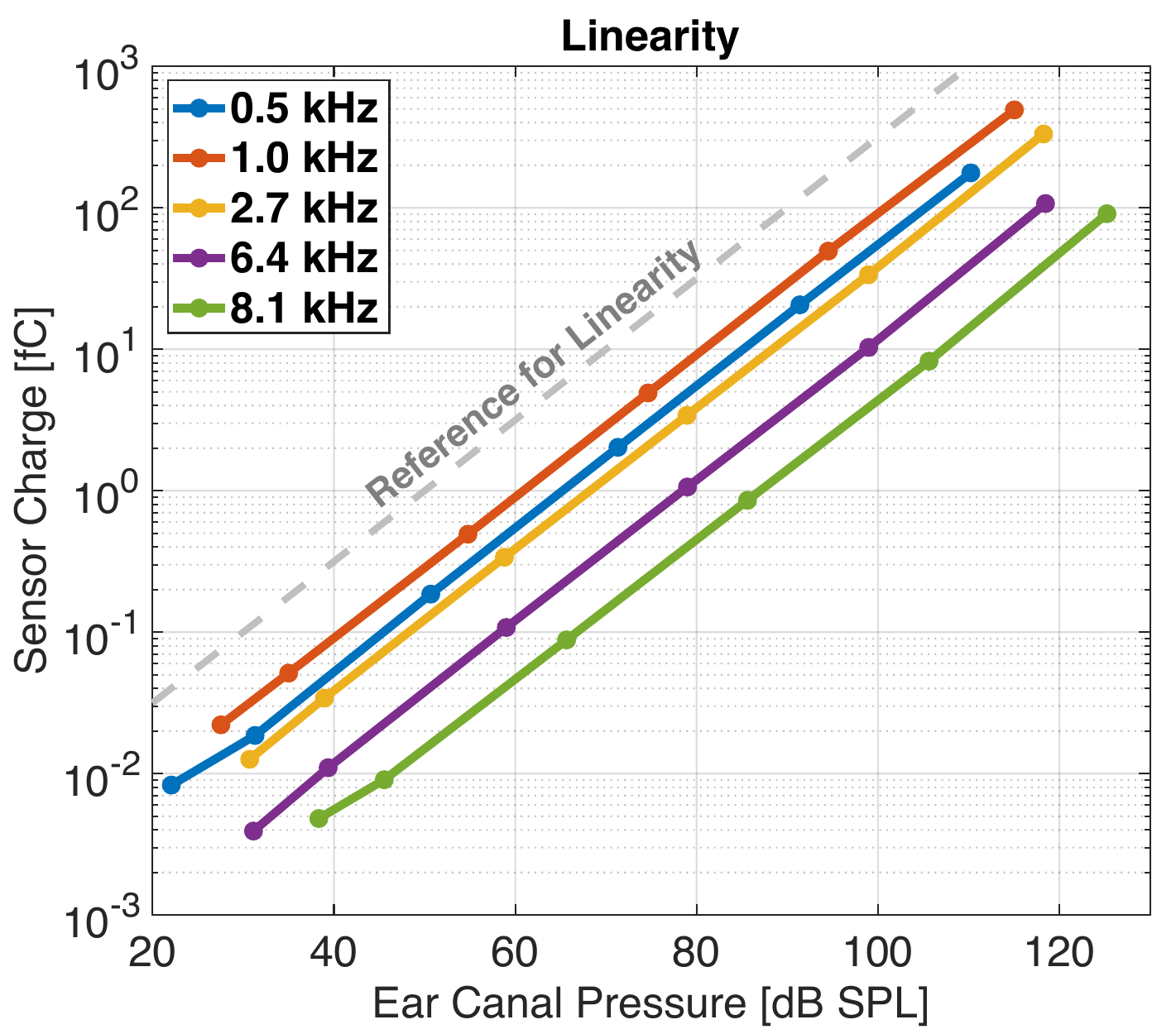}
  \caption{No significant nonlinearity across \SI{90}{dB} of dynamic range.}
  \label{fig:lin}
\end{figure}

%\marginpar{\tiny include static offset and insertion depth}

A hearing device should ideally have a flat frequency response from \SI{100}{Hz} to \SI{4}{kHz}, as this is the frequency range of human speech~\cite{device_review}. The UmboMic apparatus performs well between \SI{100}{Hz} to \SI{7}{kHz}, with the frequency response determined mostly by the middle ear impedance; Figure~\ref{fig:freqresp} shows the frequency response of the UmboMic apparatus normalized to ear canal pressure (where the responses were confirmed to be in the linear region). 
Below about \SI{1}{kHz}, the middle ear is spring-like and the frequency response of the UmboMic apparatus is flat.
Above 5~to~\SI{6}{kHz} the mass of the eardrum and ossicles dominates, causing umbo motion and thus sensor output to start decreasing.

The pinna and ear canal act like a horn and provide up to \SI{20}{dB} of pressure gain between \SI{2}{kHz}~and~\SI{6}{kHz} \cite{Shaw1974}. The cadaveric specimens that we work with no longer have the pinna attached, but we can use the known transfer function of the pinna~\cite{pinna_transfer} to simulate the pressure gain from the outer ear and extrapolate free field data. The dotted line in Figure~\ref{fig:freqresp} shows the result from including the pinna. The grey line shows the noise floor of the sensor in units of \SI{}{fC}.

We can compare the UmboMic apparatus to existing microphones through equivalent input noise (EIN), the level of acoustic noise due to the intrinsic electrical noise of the system.
We compute EIN by dividing noise floor by sensitivity and normalizing to 1/3-octave bandwidth. The EIN is a critical metric as it is related to the lowest sound that the microphone can sense.
Figure~\ref{fig:ein} shows the UmboMic apparatus EIN compared to that of a commercial hearing aid microphone, the Knowles EK3103.
We additionally simulate the EIN when including pressure gain from the outer ear. When accounting for this pressure gain, the UmboMic apparatus is competitive---we measured an A--weighted EIN of \SI{32.3}{dB}~SPL from \SI{100}{Hz}~to~\SI{7}{kHz}.
Our Knowles EK3103 reference microphone measured \SI{33.8}{dB}~SPL over the same frequency range.

Dynamic range and linearity are significant concerns for hearing aid microphones.
A frequency-domain plot of the UmboMic apparatus response to a \SI{1}{kHz} stimulus is shown in Figure~\ref{fig:harmdist}, demonstrating less than \SI{0.1}{\percent} harmonic distortion at \SI{94.5}{dB}~SPL at the eardrum.
At a \SI{114.5}{dB}~SPL stimulus level, harmonic distortion was measured to be less than \SI{1}{\percent}. Additionally, Figure~\ref{fig:lin} shows that the UmboMic apparatus is linear across at least \SI{80}{dB} of sound stimulus level.

The UmboMic apparatus also effectively rejects EMI from common sources like switched mode power supplies and \SI{60}{Hz} mains hum.
Our measured ``EMI capacitance'' was approximately \SI{0.6}{fF}, which represents an improvement of roughly \SI{54}{dB} from our lab's older single-ended unshielded designs \cite{cary2022}.
We also measured minimal interference from \SI{60}{Hz} mains power and harmonics and minimal electrical coupling between the test speaker and the sensor.
% In Figure~\ref{fig:ein} peaks at 120, 180, and \SI{300}{Hz} corresponding to the second, third, and fifth harmonics of mains frequency can be seen in the Knowles EK3103 noise floor---these peaks are absent in the cantilever--mic's noise floor.
%, equivalent to an unshielded \SI{11}{\um} sphere.
%
%\marginpar{\tiny this comparison is not too helpful. can we be more precise/ meaningful?}
%
%\textcolor{red}{In contrast, we measured an EMI capacitance five hundred times higher for lab's older single-ended unshielded designs.}

\section{Conclusion}

%\marginpar{\tiny should summarize the key metrics and comparisons to commercial devices}
%
%The microphone performance described here demonstrates the electromechanical feasibility of a %microphone built from a PVDF cantilever.
%Future work is required to study  biological compatibility and longevity.
%These considerations include materials selection, device packaging, encapsulation, and surgical hardware to securely anchor the cantilever in place.
%One must also consider electrical power consumption and the power delivery system to the implanted device.

A totally-implantable cochlear implant would significantly improve the lives of users. The microphone component is one of the largest roadblocks to internalizing the entire system. Here, we present the UmboMic, a proof-of-concept prototype of a PVDF-based microphone that senses the motion of the umbo. We demonstrate that PVDF can work well as a sensing material if designed as double-layered and paired with a very low-noise differential amplifier. When considering the effect of the pinna on performance, the UmboMic apparatus achieves an EIN of \SI{32.3}{dB}~SPL over the frequency range \SI{100}{Hz}~to~\SI{7}{kHz}---competitive with conventional hearing aid microphones. Furthermore, the UmboMic apparatus has a flat frequency response to within $\sim$\SI{10}{dB} from approximately \SI{100}{Hz} to \SI{6}{kHz}, low harmonic distortion, excellent linearity, and good shielding against EMI. 

 Our prototype demonstrates the feasibility of a PVDF-based microphone. Our future goals are to re-engineer the UmboMic sensor out of biocompatible materials. We plan to use conductors such as titanium or platinum for the patterned electrodes, and replace the flex PCB with a version made in-house from biocompatible materials. Additionally, we must consider device packaging, power system, and surgical hardware to securely hold the UmboMic apparatus in place. While these engineering challenges are substantial, our results demonstrate a suitable design concept for an implantable microphone which is competitive in performance to conventional hearing-aid microphones.

\section{Acknowledgements}

Kurt Broderick's (MIT.nano) and Dave Terry's (MIT.nano) expertise were instrumental in designing the UmboMic's fabrication process.
Many thanks to Yew Song Cheng (MEE, UCSF) for helping carry out temporal bone experiments at Mass.\ Eye and Ear.

\bibliographystyle{unsrt}
\bibliography{references, chargeamp}

\end{document}